\documentclass[a4paper,twoside,10pt]{article}
\usepackage[T1]{fontenc}
\usepackage[latin1]{inputenc}
\usepackage{amsmath, amssymb, latexsym, amsthm, mathrsfs}
\usepackage{graphicx}
\usepackage{ulem}
\usepackage{dsfont}
\input{epsf}
\usepackage{hyperref}

\date{}
\def\XXint#1#2#3{{\setbox0=\hbox{$#1{#2#3}{\int}$}
\vcenter{\hbox{$#2#3$}}\kern-.5\wd0}}

\renewcommand{\epsilon}{\varepsilon}
\renewcommand{\phi}{\varphi}

\theoremstyle{plain}

\theoremstyle{remark}

\usepackage{listings}
\usepackage{color}
\title{Numerically more stable computation of the p-values for the two-sample Kolmogorov-Smirnov test}

\author{Thomas Viehmann
\thanks{MathInf GmbH, tv@mathinf.eu}
}
\date{MathInf Technical Report 2021-1, February 2021}
\begin{document}

\maketitle

\begin{abstract}
  The two-sample Kolmogorov-Smirnov test is a widely used statistical test for detecting whether two samples are likely to come from the same distribution.
  Implementations typically recur on an article of Hodges from 1957.
  The advances in computation speed make it feasible to compute exact p-values for a much larger range of problem sizes,
  but these run into numerical stability problems from floating point operations. We provide a simple transformation of the defining recurrence for the two-side two-sample KS test that avoids this.
\end{abstract}

The Kolmogorov-Smirnov two sample test (KS test) is perhaps the go-to statistical statistical test of whether two samples originate form the same distribution.

To make things precise, we consider samples $x_1, \dots, x_m \in \mathbb{R}$ and $y_1, \dots, y_n \in \mathbb{R}$ drawn from two continuous distribution functions $F$ and $G$, respectively.
We form the empirical distribution functions $F_m(x) := |\{x_i : x_i \leq x \}|/m$ and $G_n(x) = |\{y_i : y_i \leq x \}|/n$. The KS test then tests the null hypothesis $F = G$.

The test-statistic is computed from the empirical distributions as

\[
  D = sup_{x} |F_m(x) - G_n(x)|.
\]
Following the usual notation, we write the supremum even though in the cases we consider, it is actually a maximum.
Working with only the ranks, the test does not make assumptions on the distributions themselves. To operationalize the test, we need to compute p-values.

Smirnov famously gave the asymptotic formula that if $m, n \rightarrow \infty$ such that $n/m \rightarrow q \in \mathbb{R}$,
$$
P_2 = Prob\left[ \sqrt{\frac{mn}{m+n}} D \geq x \right] \rightarrow 1 - K(x) = 2 \sum_{k=1}^{\infty} (-1)^{k-1} \exp (-2k^2x^2),
$$
where $K$ is the cumulative distribution function of the Kolmogorov distribution. We use Hodges' name $P_2$ for the two-sided problem.

However, as noted by Hodges, it is unclear how well they work in practice: After studying the problem numerically for $m=12$, $n=13,\dots,18$ he writes ``The
Smirnov approximation is seen to be highly inaccurate for values of m and n which are already large enough for direct computations to be arduous.'' The amount of computation that is possible with ease has dramatically increased since the 1950s and so it is natural to revisit direct computation.

\section{Direct computation of p-values}

The key observation for the computation of p-values is that we can compute the distribution of `D` under the null hypothesis in purely combinatorical terms.
Under the assumptions, all values are distinct with probability 1. We may order the joint sequence of the $x_i$ and $y_j$. We ignore the indices and write an
$x$ in positions where a $x_i$ occurs, and $y$ where $y_i$ occurs, obtaining a random sequence of $m$ $x$es and $n$ $y$s.
Under the null hypothesis, all possible $\binom{m+n}{m}$ distinct sequences have equal probability.

We may map these to paths $P$ between $(0, 0)$ and $(1, 1)$  where for each $x$ we move to the right by $1/m$ and for each $y$ we move up by $1/n$.
Given a path $P$, the statistic $D$ is the maximum $\sup_{(x,y) \in P} |x-y|$.
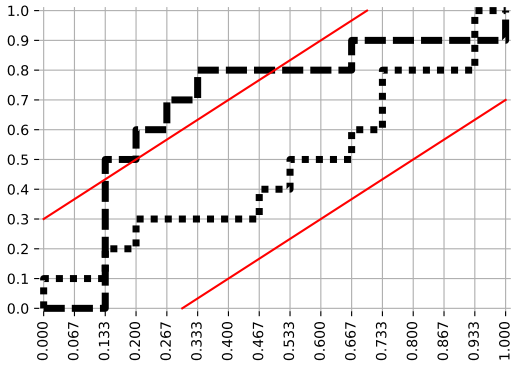
\includegraphics[width=8cm]{two-paths.pdf}

To compute the probability that $D \leq d$ we thus need to count the $(m,n)$-step paths such that all grid points are in the corridor $|x-y| < d$ and devide this number $A_{m,n}$ by the total number of paths $\binom{m+n}{n}$.

The classical iteration (\cite{Hodges}) is

\[
  A_{i,j} =
  \begin{cases}
    0 & \text{if } |i/m-j/n| \geq d, \\
    1 & \text{if } |i/m-j/n| < d \text{ and } (i = 0 \text{ or } j = 0), \\
    A_{i-1, j} + A_{i, j-1} & \text{otherwise.}
  \end{cases}
\]
As we are iterating over the nodes inside the corridor (because we will not compute the zero values except perhaps from rounding up to a fixed number of nodes), this is called the \textit{inside method}. Hodges notes that the number of additions increases as $n^{\frac{3}{2}}$ as problem sizes increases and that operands increase exponentially. With today's computational resources, we are only mildly concerned about the number of operations. However, even in today's software packages the size of operands translates into difficulties with numerical accuracy. Taking SciPy \cite{SciPy} as an example, when the probability $P_2$ is computed, they first compute $1-P_2$ using the inside method (with appropriate stabilization) but it can happen that $1-P_2$ is larger than the smallest number smaller than $1$ representable by the floating point numbers used.

We therefore propose to use a different scheme to compute $P_2$ that avoids computing $1-P_2$. We define
\[
  C_{i, j} = 1 - A_{i, j} / \binom{i + j}{i}
\]
as the proportion of paths from $(0,0)$ to the grid point $(i/m, j/n)$ that do not stay inside the $d$-corridor. At the top right corner, $C(m,n) = P_2$. For $(i, j)$ inside the corridor we have
\begin{align*}
  C_{i, j} &= 1 - A_{i, j} / \binom{i + j}{i} \\
   &= 1 - (A_{i-1, j} + A_{i, j-1}) / \binom{i + j}{i} \\
   &= 1 - \left((1 - C_{i-1, j}) \binom{i+j-1}{i-1} + (1 - C_{i, j-1}) \binom{i+j-1}{i}\right) / \binom{i + j}{i} \\
   &= 1 - \left((1 - C_{i-1, j}) \frac{i}{i+j} + (1 - C_{i, j-1}) \frac{j}{i+j}\right) \\
   &= C_{i-1, j} \frac{i}{i+j} + C_{i, j-1} \frac{j}{i+j}.
\end{align*}
Combined with the conditions for when $C_{i,j}$ is $0$ or $1$ we get
\[
C_{i,j} =  \begin{cases}
    1 & \text{if } |i/m-j/n| \geq d, \\
    0 & \text{if } |i/m-j/n| < d \text{ and } (i = 0 \text{ or } j = 0), \\
    C_{i-1, j} \frac{i}{i+j}  + C_{i, j-1} \frac{j}{i+j} & \text{otherwise,}
  \end{cases}
\]
a structurally very simple recursion.

Note that in the corridor, $C_{i,j}$ is a weighted average of its neighbours, so we may expect favourable error propagation. When the value decreases in the iteration, it typically is because one of the predecessor is $0$ or smaller than the other, i.e. through scaling rather than cancellation.

\section{Implementation}

We implemented the recursion on $C$ in Python, Numba\cite{Numba} and C++. The (naively implemented) computation is considerably slower (approximately 4x) than the path-counting algorithm
implemented by SciPy (when both are either run in Python or through the numba compiler), likely at least in part due to the additional multiplications and divisions in the retursion for $C$. It is possible to exploit GPU-type parallelism using the scanning pattern also used by operations like cumulative sum.

As noted by Hodges, it suffices to keep track of the values in a $d$-corridor along the diagonal, so a naive Python implementation (that also can be compiled to native code using Numba).
There are other possibilities for speedup, not lest that due to the rotational symmetry outlined by Hodges, which we do not implement.

\begin{lstlisting}[caption=Stabilized Inner Method]
def compute_p2(n, m, d):
    size = int(2*m*d+2)
    lastrow, row = numpy.zeros((2, size), dtype=numpy.float64)
    last_start_j = 0
    for i in range(n + 1):
        start_j = max(int(m * (i/n + d)) + 1-size, 0)
        lastrow, row = row, lastrow
        val = 0.0
        for jj in range(size):
            j = jj + start_j
            dist = i/n - j/m
            if dist > d or dist < -d:
                val = 1.0
            elif i == 0 or j == 0:
                val = 0.0
            elif jj + start_j - last_start_j >= size:
                val = (i + val * j) / (i + j)
            else:
                val = (lastrow[jj + start_j - last_start_j] * i + val * j) / (i + j)
            row[jj] = val
        jjmax = min(size, m + 1 - start_j)
        last_start_j = start_j
    return row[m - start_j]
\end{lstlisting}

\section{Conclusion}

We have outlined an alternative iteration for computing p-values for the 2-sample KS test for the inside method that exibits better numerical stability.
It would be desirable to also obtain an analogue for the outside method.

\end{document}